# Understanding the image contrast of material boundaries in IR nanoscopy reaching 5 nm spatial resolution

*Stefan Mastel[1], Alexander A. Govyadinov[1], Curdin Maissen[1], Andrey Chuvilin[1,2], Andreas Berger[1], Rainer Hillenbrand*[*2,3]*

[1]CIC nanoGUNE, 20018 Donostia-San Sebastián, Spain
[2]IKERBASQUE, Basque Foundation for Science, 48013 Bilbao, Spain
[3]CIC nanoGUNE and UPV/EHU, 20018 Donostia-San Sebastián, Spain

[*]*Corresponding author: r.hillenbrand@nanogune.eu*

**Scattering-type scanning near-field optical microscopy (s-SNOM) allows for nanoscale-resolved Infrared (IR) and Terahertz (THz) imaging, and thus has manifold applications ranging from materials to biosciences. However, a quantitatively accurate understanding of image contrast formation at materials boundaries, and thus spatial resolution is a surprisingly unexplored terrain. Here we introduce the write/read head of a commercial hard disk drive (HDD) as a most suitable test sample for fundamental studies, given its well-defined sharp material boundaries perpendicular to its ultra-smooth surface. We obtain unprecedented and unexpected insights into the s-SNOM image formation process, free of topography-induced artifacts that often mask and artificially modify the pure near-field optical contrast. Across metal-dielectric boundaries, we observe non-point-symmetric line profiles for both IR and THz illumination, which are fully corroborated by numerical simulations. We explain our findings by a sample-dependent confinement and screening of the near fields at the tip apex, which will be of crucial importance for an accurate understanding and proper interpretation of high-resolution s-SNOM images of nanocomposite materials. We also demonstrate that with ultra-sharp tungsten tips the apparent width (and thus resolution) of sharp material boundaries can be reduced to about 5 nm.**

Scattering-type scanning Near-field Optical Microscopy (s-SNOM)[1] is a scanning probe technique for visible, infrared, and terahertz imaging and spectroscopy with nanoscale spatial resolution. It has proven large application potential ranging from materials characterization[2,3] to biosciences.[4,5] In s-SNOM, a metalized atomic force microscope (AFM) tip is illuminated with p-polarized light. The tip acts as an antenna and concentrates the illumination at its apex to a near-field spot on the scale of the apex radius. When brought into close proximity to a sample, the near field interacts with the sample and modifies the tip-scattered field[6]. By recording the tip-scattered field while scanning the sample, a near-field image is obtained. It is generally accepted that essentially the tip's apex radius determines the achievable resolution, which is typically in the range of a few tens of nanometers.[7,8] Although the resolution is a key parameter in s-SNOM - as in any other microscopy technique - it has been barely studied in detail experimentally.

The spatial resolution in microscopy is often evaluated by measuring the width of a typically point-symmetric line profile across the sharp boundary between two different materials.[9-11] Such line profiles can be considered as the so-called Edge Response Function (ERF). The characteristic width $w$ of the ERF can be determined via its derivative, which is also known as the Line Spread Function (LSF). The LSF represents the image of a line-like object and is typically a bell-shaped symmetric function centered at the material boundary. The width of the LSF determines according to a specific criterion such as the Rayleigh or Sparrow.[12]

In s-SNOM experiments, $w$ (often interpreted as the spatial resolution in analogy to other microscopy techniques), is typically measured directly in line profile recorded across the boundary[13-16] or via its derivative[17]. Values as small as $w = 10$ to $40$ nm (evaluated using different criteria) have been reported for a broad spectral range extending from visible to terahertz frequencies.[13,14,18] However, the boundary between two different materials typically exhibits a step in topography, which challenges the reliable evaluation of $w$ due to tip-sample convolution,[19-22] potentially resulting in a large over- or underestimation. To tackle this problem, a sample with a well-defined sharp material boundary but without topographic features is highly desired.[19,20]

Here we introduce the read/write head of a hard disk drive (HDD) as a truly topography-free resolution test sample, exhibiting nanoscale-defined metal-dielectric boundaries perpendicular to its ultra-smooth surface. It serves as an analogue to the knife-edge test target[10,11] in classical optical microscopy and allows for detailed analysis of the s-SNOM image contrast with metal tips of apex radii down to 3 nm. We demonstrate that with these tips the ERF width (evaluated as full width half maximum of the corresponding LSF) $w$ can be smaller than 5 nm. We further find, surprisingly, that the derivative of the ERF in s-SNOM is generally an asymmetric function. Its width depends on the side of the material boundary where it is evaluated. On the metal side, we find an unexpectedly short near-field probing range that can be one order of magnitude below the tip apex diameter, which we explain by screening of the tip's near field by a metal sample. We corroborate our results by numerical simulations and discuss the implications of our findings for the interpretation of s-SNOM images in general.

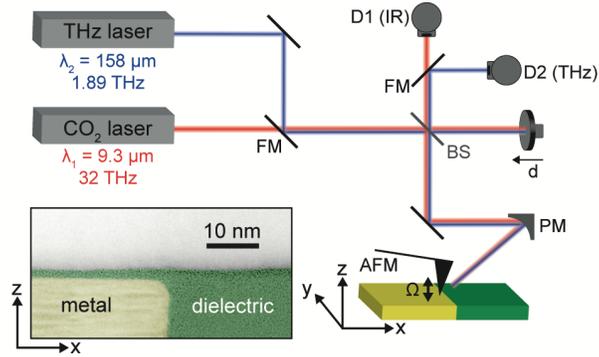

**Figure 1:** Schematics of the THz and IR s-SNOM setup. AFM, atomic force microscope; FM, flip mirror; BS, beam splitter; PM, parabolic mirror; D1, IR detector; D2, THz detector. The inset shows a STEM image of a cross section of our sample, which consists of the edge of a magnetic shield structure in a read/write HDD head.

Figure 1 shows the experimental setup and the HDD read/write-head sample. For measurements we utilized a commercial s-SNOM (Neaspec GmbH). The tip was illuminated by either a $CO_2$ ($\lambda_1 = 9.3$ μm) or a THz ($\lambda_2 = 158$ μm) laser beam with the polarization plane parallel to the tips axis. The tip acts as an antenna and concentrates the incoming radiation at the tip apex. In close proximity to a sample, the near fields interact with a sample and modify the tip-scattered field. The tip-scattered light is recorded by detector D1 (IR) or D2 (THz), and contains information about the local optical properties of the sample. An interferometric detection scheme, operated in synthetic optical holography (SOH) mode,[23] enables the recording of both amplitude s and phase ϕ images. For background suppression, the tip is oscillated vertically at a frequency Ω and the tip-scattered signal is demodulated at higher harmonics n of the cantilever oscillation frequency Ω, yielding background-free near-field amplitude $s_n$ and phase $\phi_n$ images.

For evaluating the resolution of the setup employing different tips, we use the read/write head of a commercial HDD as resolution test sample, and more specifically the edge of one of its magnetic shield structures. The lower left inset in Fig. 1 shows a false color Scanning Transmission Electron Microscopy (STEM) image of a cross section of the sample. The contrast in the image lets us recognize sharply separated areas of metal (marked yellow) and dielectric material (marked green). According to the manufacturer of the HDD,[24] the metal is Permalloy (Fe/Ni 20/80), and the dielectric is $Al_2O_3$. Further, we observe in the STEM image a dielectric capping layer of around 1.5 nm covering the metal, and thus also the material boundary. Most importantly for s-SNOM imaging, the STEM image shows that the sample surface is smooth down to the sub-nm scale, even in the sample area, where the material changes abruptly.

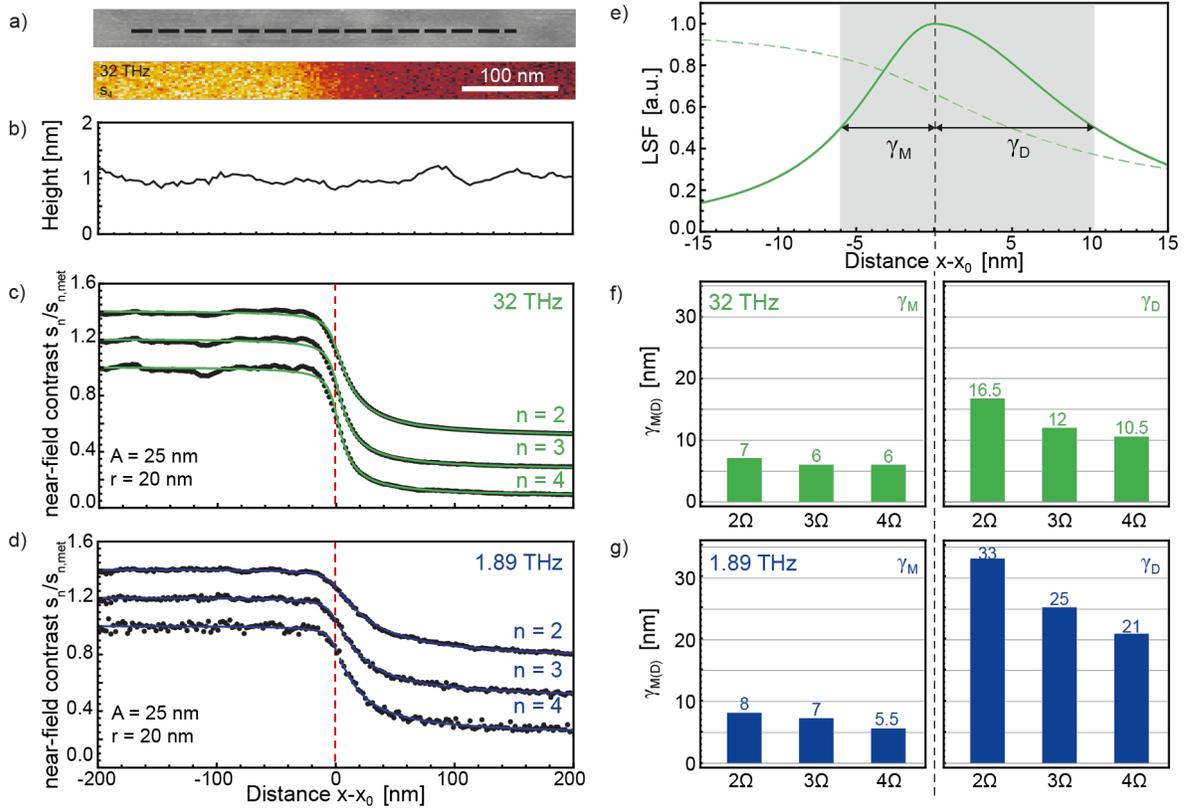

**Figure 2:** s-SNOM measurements on the resolution test sample. a) AFM topography and IR s-SNOM amplitude $s_4$ ($\lambda$ = 9.3 µm) images of sample. b) Topography line profile extracted along the dashed line in a). c,d) Measured IR and THz near-field amplitude contrast $s_n/s_{n,met}$ line profiles (average of 20) for harmonics n = 2 to 4 (black dots), and their respective fits using the integral of an asymmetric Lorentzian as described in text (green/blue lines). Tapping amplitude A = 25 nm, tip radius r = 20 nm. The curves are vertically offset for better visibility. e) Derivative (solid line) of the fit of the $s_4$ line profile (dashed line) taken from panel d). f,g) HWHM $\gamma_{M(D)}$ of the derivatives of line profile fits in c) and d) on the metal and dielectric side, respectively.

Figure 2 shows the s-SNOM imaging results of the material boundary. As near-field probe we employed an 80 µm long Pt/Ir tip (Rocky Mountain Nanotechnology, RM), operated at A = 25 nm tapping amplitude. We used long RM tips rather than standard cantilevered Pt/Ir-coated AFM tips (around 10 µm long) because of their better performance as near-field probes in the THz spectral range.[25,26] The RM tip radius of r = 23 nm is comparable to the standard metal-coated tips utilized in s-SNOM. We first recorded a topography image of the sample (Fig. 2a), from which we extracted a line profile (Fig. 2b) along the black dashed line. The line profile shows a maximum topography variation of 4 Ångstrom, which confirms the flatness of the sample. Simultaneously with topography, we recorded IR (32 THz) and THz (1.89 THz) s-SNOM amplitude images from

$s_2$ to $s_4$. As an example, we show in Fig. 2a the IR $s_4$ image. We observe two regions with high and low near-field amplitude signal, which lets us recognize the metal and dielectric material, respectively.[13,15]

To analyze the s-SNOM signal transition across the material boundary, we averaged 20 line profiles for each of the IR and THz images $s_2$ to $s_4$ (see Methods). The averaging ensures an accurate measurement of the apparent width of the material boundary, as individual line profiles can exhibit an untypically small or large width due to noise (see supplementary information). The averaged line profiles (black dots) are shown in Fig. 2c and d. In agreement with former observations, we observe that (i) the near-field contrast (i.e. the ratio between the near-field signal on metal and on dielectric material) increases for increasing demodulation orders n[27-29] and (ii) the contrast is higher for the IR than for THz. In order to better visualize the effect of demodulation order and on the near-field contrast, we show in the Supporting Information S4 the same line profiles as in Fig. 2c and d, but not vertically offset. The difference between the IR and THz material contrast can be attributed to frequency-dependent dielectric permittivities of the sample. Most important, and not having been recognized in previous s-SNOM experiments, the line profiles in Fig. 2c and d are *not* point-symmetric, which we will study and discuss in the following.

The *asymmetric* line profiles require a careful analysis in order to properly interpret the s-SNOM contrast at material boundaries. As a first step towards this goal, we approximate the line profiles by the empirically found fit function:

$$\Theta(x) = \begin{cases} \pi^{-1} f_M \, \text{Arctan}\left(\frac{x-x_0}{\gamma_M}\right) + b & \text{for } x < x_0 \\ \pi^{-1} f_D \, \text{ArcTan}\left(\frac{x-x_0}{\gamma_D}\right) + b & \text{for } x \geq x_0 \end{cases},$$

with fit parameters $x_0$ (interface position) and $b$ (vertical offset). To account for the asymmetry of the line profiles, the fit parameters $f_{M,D}$ and $\gamma_{M,D}$ assume different values for the metal ($x < x_0$) and the dielectric ($x \geq x_0$) sides; the continuity of $\Theta$ and its derivative across the material interface are further enforced. These fits are shown as green and blue solid curves in Fig. 2c and d, excellently matching the experimental data. In the Supporting Information S2 we show a fitting of the line profiles with a symmetric function. We find that the agreement between data and fits are much worse, showing that indeed asymmetric fitting is required to correctly analyze the experimental line profiles. We next use these fits to quantify the asymmetry of the line profile. To that end, we calculate the derivative of the fit function $\Theta(x)$, which is given by a piecewise Lorentzian (exemplarily shown in Fig. 2e for the IR $s_4$ line profile):

$$\mathcal{L}(x) = \begin{cases} \frac{f_M}{\pi} \frac{\gamma_M}{(x-x_0)^2 + \gamma_M^2} & \text{for } x < x_0 \\ \frac{f_D}{\pi} \frac{\gamma_D}{(x-x_0)^2 + \gamma_D^2} & \text{for } x \geq x_0 \end{cases},$$

with different half width at half maxima (HWHM) $\gamma_{M(D)}$ for the metal and the dielectric sides. Note that in the context of this work, we call this derivative the Line Spread Function (LSF) in analogy to the general concepts of resolution in classical optical microscopy as described in the introduction. The bar diagrams in Figs. 2f and g summarize the different values for $\gamma$ of the IR and THz measurements for demodulations orders n = 2, 3, 4. We find that the $\gamma_D$ are about three to four times larger than the $\gamma_M$, quantifying the significant asymmetry of the line profiles. The total material boundary width $w$, defined as $w = \gamma_M + \gamma_D$, decreases from 23.5 nm to 16.5 nm (IR line profile) and from 41 nm to 26.5 nm (THz line profile) when the demodulation order increases from n = 2 to an = 4 (Fig. 2f,g). The sharpening of the material boundary by higher-harmonic demodulation and the values for $w$ agree well with previous studies,[28,30,31] which, however, did not recognize the asymmetry of the profiles. Our quantitative analysis further shows that the material boundary is located not exactly central to the signal transition (see further discussion below), which is critical when a precise localization of material boundary form s-SNOM profiles is desired. The analysis also shows that a significant near-field signal tail into one material does not necessarily indicate unidirectional material gradient, for example caused by unidirectional material diffusion. Our results clearly show that asymmetric line profiles with substantial levels of asymmetry can occur at well-defined sharp material boundaries, a fact that seems to be intrinsic to the near-field interaction and probing process. We will elucidate this phenomenon below.

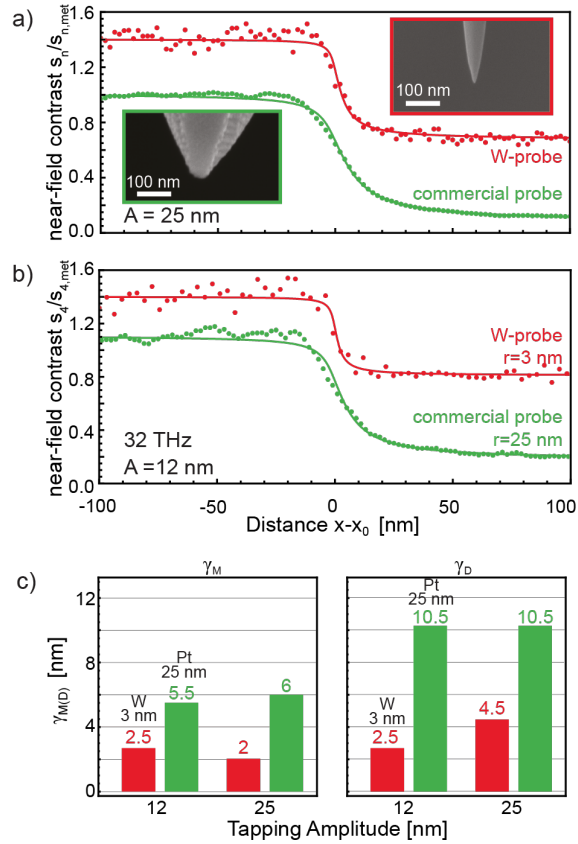

**Figure 3:** s-SNOM resolution test measurements for different probe sizes at 32 THz. a) Near-field amplitude contrast $s_4/s_{4,met}$ line profiles recorded with ultra-sharp W tip (r = 3 nm) (red dots) and commercial Pt/Ir tip (r = 25 nm) (green dots, same data as in Fig. 2c) at A = 25 nm tapping amplitude. The green and red solid lines show the respective fits on the data. The upper right and lower left inset shows an SEM image of the W-tip and Pt/IR-tip, respectively. The curves are vertically offset for improved representation. b) Near-field amplitude contrast $s_4/s_{4,met}$ line profiles and their respective fits recorded with ultra-sharp W tip (r = 3 nm) (red) and commercial Pt/Ir tip (r = 25 nm) (green) at A = 12 nm tapping amplitude. c) $\gamma_{M(D)}$ evaluated for the line profiles recorded with the W- and Pt/Ir-tips at A = 25 nm and A = 12 nm tapping amplitudes.

To reduce the perceived width *w* of the material boundary in the s-SNOM line profile, i.e. to increase the spatial resolution, we employed Focused Ion Beam (FIB) machining to fabricate Tungsten (W) tips with a reduced tip radius of only r = 3 nm (see upper right SEM image in Fig. 3).[8,25,32,33] Utilizing the ultra-sharp full-metal W probes, we recorded line profiles across the material boundary at 32 THz illumination and a tapping amplitude of A = 25 nm. The red dots in Fig. 3a show the $s_4$ line profile (average of 50 profiles, see Methods) and the corresponding fit (red curve). For comparison we show the line profile obtained with the Pt/Ir probe (green; same data and fit as in Fig. 2c). By measuring $\gamma_{M(D)}$ for both line profiles (summarized in Fig. 3c), we find that $w$ is reduced by more than a factor of two when the W-tip is used. The improvement, however, is surprisingly small, considering that the tip radius of the W tip is around eight times smaller than that of the Pt/Ir tip. We attribute this finding to the relatively large tapping amplitude of A = 25 nm, which is comparable to the radius of the Pt/Ir tip (r = 25 nm) but much larger than the radius of the W tip (r = 3 nm). According to previous studies,[28,34] the width $w$ can be improved by reducing the tapping amplitude. We thus recorded a line profile using both the Pt/Ir and W tip with a reduced tapping amplitude of A = 12 nm (green and red data in Fig. 3b, respectively). The resulting values for $\gamma_{M(D)}$ are shown in Fig. 3c. For the Pt/Ir tip, we measure $w$ = 16.5 nm, which is comparable to that of the line profile at larger tapping amplitude A = 25. For the W-tip the width $w$ of the material boundary decreases further, to about 5 nm, which clearly demonstrates that ultra-sharp metal tips can push the s-SNOM resolution well below 10 nm. We note that this reduction is mainly caused by the reduced $\gamma_D$ of the LSF on the dielectric side of the material boundary. On the metal side, the 1.5 nm-thick dielectric capping layer makes the metal/dielectric boundary a subsurface object (SEM image; Fig. 1), for which the resolution is well known to be diminished compared to objects directly at the surface.[28,35] It also has to be noted that numerous experiments reliably reveal a decrease of the $s_n$-signal with decreasing tip diameter, which requires averaging of several line profiles to obtain sufficiently high signal to noise ratios. We attribute this behavior to the stronger localization of the near field for sharper tip apices and thus the reduction of the sample volume participating in the near-field interaction with the tip, which is not compensated by the increased field enhancement at sharper tip apices.

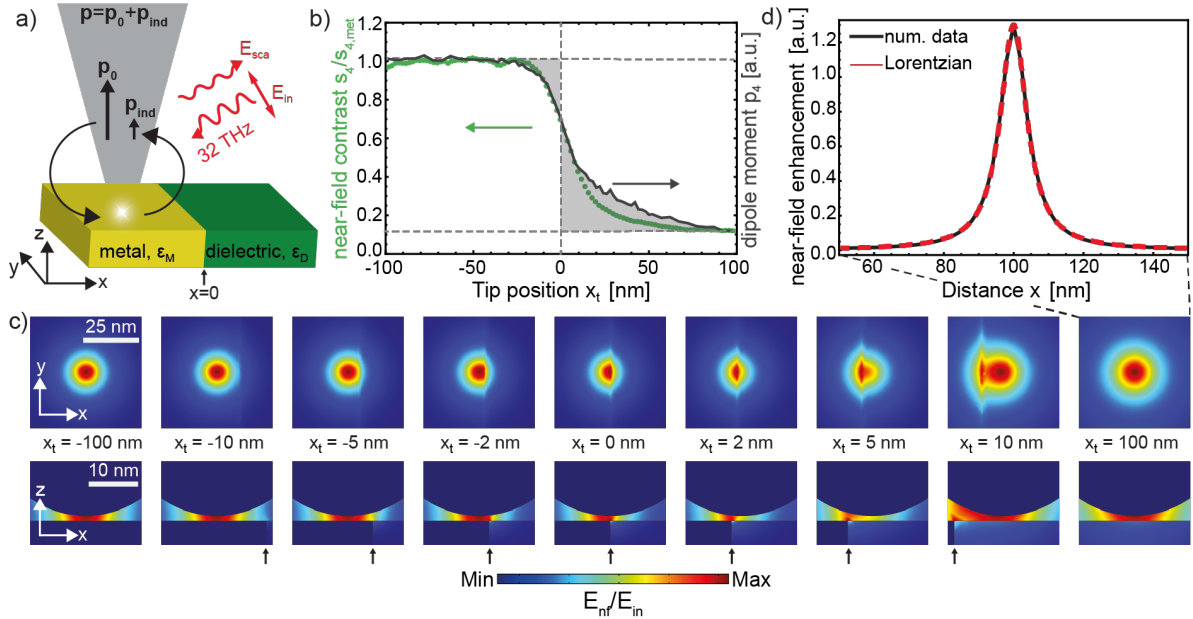

**Figure 4:** Numerical simulation of IR s-SNOM line profiles. a) Sketch of the geometry. A tip with apex radius r = 25 nm and length 8 μm is placed above a sample consisting of metal on the left (x < 0 nm) and dielectric material on the right (x > 0 nm) side. The material boundary is at x = 0. b) Simulated (blue curve) and measured (green dots, same data as in Fig. 2c) s-SNOM amplitude signal contrast $s_4(x_t)/s_{4,met}(x_t)$ for a tapping amplitude A = 25 nm for different tip positions $x_t$ relative to the material boundary. c) Electric near-field distribution below the tip apex for different tip positions $x_t$ in the xy-plane (z = 0 nm) and xz-plane (y = 0 nm) for tip-sample separation of 1 nm. The arrow marks the material boundary. d) Profile of the simulated electric near-field distribution along the x-axis when the tip is placed above the dielectric material (black curve) ($x_t$ = 100 nm). Fit of a Lorentzian function (red curve) to the simulated near-field profile.

In Figure 4 we show results of a numerical study, which aims at corroborating and understanding the asymmetry of the s-SNOM line profiles observed in our experimental study (Figures 2 and 3). We performed numerical full-wave simulations of the s-SNOM imaging process using the commercial software package Comsol. A conical tip of 8 μm length and apex radius $r$ = 25 nm length is placed above a sample modeled by metallic permittivity of $\varepsilon_M$ = -1200 + 750$i$ on the left side ($x$ < 0 nm) and a dielectric material of $\varepsilon_D$ = 1.05 + 0.19$i$ (Al$_2$O$_3$)[36] on the right side ($x$ > 0 nm) of the material boundary at $x$ = 0 nm (see illustration in Fig. 4a). We assume a p-polarized plane wave illumination (electric field $E_{in}$) at 32 THz at an angle of $\alpha$ = 60° relative to the tip axis, as in our s-SNOM. The tip-scattered electric field $E_{sca}$ is proportional to the complex-valued dipole moment $P$, calculated numerically according to[37]

$$E_{sca} \propto P = \int \sigma(\mathbf{r}) \mathbf{r}\, d\mathbf{r},$$

where $\sigma(\mathbf{r})$ is the surface charge density, $\mathbf{r}$ is the radius vector, and the integral is carried out over the whole tip surface. Note that $P$ can be considered as the sum of the tip's dipole moment $p_0$ arising from the polarization induced by the incoming radiation $E_{in}$ and the induced dipole moment $p_{ind}$ originating from the tip's near-field interaction with the sample, the latter yielding the s-SNOM signal. To simulate the measured s-SNOM signal we have to take into account that the tip is oscillating, and the detector signal is demodulated at higher harmonics of $n\Omega$. Accordingly, we first calculate the scattered field, $E_{sca}(z_t)$, as a function of tip height $z_t$ above the sample. Assuming a vertical sinusoidal motion of the tip with frequency $\Omega$ and tapping amplitude $A = 24.5$ nm, we calculate the time evolution of the detector signal $E_{sca}(z_t(t))$ with $z_t(t) = 0.5 + A/2 * (1 + \cos(\Omega t))$. The n-th Fourier coefficient of $E_{sca}(z_t(t))$ is then the mathematical analogue of the complex-valued s-SNOM signal $s_n e^{i\phi_n}$. By calculating $s_n$ as a function of tip position $x_t$, we obtain the simulated line profile $s_n(x_t)$ across the material boundary. The blue curve in Fig. 4b shows the result obtained for demodulation at n = 4. For comparison, we also show the experimental line profile $s_4(x)$ (red dots, same data as in Fig. 2c). A good match between the simulated and the experimental line profiles is found after normalization of both near-field profiles to their average value on metal. We note that the model over-predicts the asymmetry slightly, essentially on the dielectric side of the material boundary. We explain this observation by differences in tip and sample geometry in experiment and simulations. For example, we simulate a perfect material boundary and a perfect conical metal tip, while in the experiment the sample´s material boundary is slightly rounded (see Fig. 1) and the tip has a more complicated (pyramidal) shape. We did not take into account the more complicated geometry in the simulation due to limited computation power. We further note that no lateral shift in x-direction was applied to the simulated data (Fig. 2c) in order to match the experimental data, which confirms the position of the material boundary found by the fitting procedure introduced in Fig. 2. Most importantly, the simulation clearly confirms the asymmetry of s-SNOM line profiles across a material boundary.

To explain the asymmetry of the line profiles, we show in Fig. 4d the calculated electric near-field distribution around the tip apex, $E_{nf}/E_{in}$, for different tip positions $x_t$. On the metal and dielectric surface, far away from the material boundary at $x_t = -100$ nm and $x_t = 100$ nm, respectively, we observe that the near-field distribution in the plane of the sample (x-y-plane) is symmetric. However, the near field confinement is markedly different, indicating a larger probing range of the tip on the dielectric side. When the tip approaches the boundary from the dielectric side, the near-field distribution is already significantly modified at $x_t = 10$ nm, revealing a near-field interaction with metal across the material boundary. Subsequently, the tip-scattered field and the s-SNOM amplitude signals $s_n$ increase. On the other hand, when the tip approaches the boundary from the metal side, a significant modification of the near-field distribution requires the tip to be closer than 5 nm to the interface ($x_t > -5$ nm). We explain this finding by the screening of the tip's near fields by the metal sample, which reduces the probing range and prevents the detection of the material boundary via the tip-scattered field for tip-boundary distances larger than 5 nm. The absence of strong near-field screening on the dielectric side thus explains the asymmetry of the s-SNOM line profiles across the boundary between metal and dielectric. In the experiment, the near-field screening by the metal is reduced due to the rounded edge of the material interface (see Fig. 1), resulting in a reduced asymmetry of the measured line profiles compared to the simulated one (Fig. 4b). We expect that the near-field screening is less important for boundaries between two materials with low dielectric contrast, which would make s-SNOM line profiles more symmetric. We finally note the electric near-field distribution below the tip apex can be well approximated by a Lorentzian function (Fig. 4d). This observation might explain why the s-SNOM line profiles can be well fitted by the integral of Lorentzian functions, but certainly further studies are required for a more comprehensive understanding. Although the presented results are discussed in the context of s-SNOM, we expect the same effect of screening to occur in images acquired by other AFM-based optical microscopy techniques, such as tip-enhanced photothermal expansion microscopy[38] and photoinduced force microscopy[39] that rely on the material profiling via tip-enhanced near fields.

In summary, we showed that the read/write head of a HDD can serve well as a topography-free test sample for fundamental s-SNOM experiments. It allowed for detailed studies of contrast, resolution, and shape of material boundaries, yielding unprecedented insights into the image contrast formation. Using tips with a standard apex diameter of about 50 nm, we find that the width *w* of a material boundary in s-SNOM images is around 20 nm, which is in agreement with former reports. However, the line profiles exhibit an asymmetry that has not been observed before, which we corroborate via numerical calculations. The asymmetry can be explained by the tip-sample near-field interaction, which has significant spatial variations across material boundaries. Particularly, we find that the near field at the tip apex is strongly screened on the metal side, which reduces the apparent width of the material boundary in s-SNOM images. We expect that a similar effect will occur at the boundary between two dielectric materials of high and low refractive index because the screening by polarization charges in high-index dielectrics is nearly as large as in metals. Considering this effect will be of critical importance for avoiding misinterpretation of asymmetric line profiles as, for example, continuous (i.e. not sharp) changes of dielectric properties caused by non-uniform doping, directional diffusion, etc. In the future, it will also be interesting to study how near-field screening affects the spatial resolution when two closely spaced objects are imaged. We further envision that near-field screening could be exploited to increase the s-SNOM resolution for molecule imaging, for example by depositing them on top of a sharp material boundary. We finally note that with custom-made ultra-sharp tips of 5 nm diameter we can reduce the apparent material boundary to about 5 nm. On the other hand, both the signal and S/N ratio decrease for sharper tips, which will require to increase the field enhancement at the apex of ultra-sharp tips, for example by engineering and optimizing the antenna performance of the tip shaft.

# Methods

## Averaging of line profiles.

The presented IR and THz line profiles in Fig. 2 recorded with the Pt/Ir are the averages of 20 single line profiles. Before averaging, we cross-correlated the line profiles for the second demodulation order n=2 in order to obtain the lateral offset between them. We then corrected for this lateral offset for each demodulation order n=2,3,4. We used the second demodulation order for finding the offsets because it provides a better SN than higher orders, which enables a higher accuracy of the cross-correlation. For the W tip line profiles presented in Fig. 3, the same procedure was applied using 50 line profiles in total.

## FIB fabrication of ultra-sharp tips

The tungsten tips were fabricated by focused ion beam (FIB) machining using a Helios 450s electron microscope (FEI, Netherlands We used standard Si atomic force microscopy (AFM) cantilevers and first made a cylindrical grove into the tip. Then, a high aspect ratio bullet was milled out of a solid tungsten wire, cut at around 12 μm length, and fitted into the cylindrical grove in the Si cantilever. The cone was attached by FIB induced deposition of silicon oxide. Details of this procedure can be found in reference 25. Finally, the tip apex was sharpened by circular ion milling along the tip axis, as described in detail in reference 32. To reach a very small tip apex diameter of 6 nm it is crucial to gradually reduce the milling current down to about 7 pA. Note that fabrication of ultra-sharp tips with radii as small as 3 nm required a hard material such as W. With Pt/Ir we achieved apex radii of about 10 nm and with Au not better than 12 nm. We assign this finding to diffusion of metal atoms under ion bombardment, which is higher for Au than for Pt/Ir and W. Further studies are needed to clarify the mechanisms involved in the tip sharpening process.

# Supporting Information

This article is accompanied by a Supporting Information document containing the following information:
- S1: Individual (not-averaged) s-SNOM line profiles recorded with Pt/Ir and W-tips.
- S2: Comparison of asymmetric vs. symmetric fit on measured line s-SNOM line profiles
- S3: Comparison of asymmetric vs. symmetric fit on s-SNOM line profiles measured with the W-tip
- S4: IR and THz line profiles without vertical offset for comparison of contrast for different demodulation orders.

# Author Contributions

S.M. and R.H. conceived the study. S.M. fabricated the tips, performed the s-SNOM experiments, fitted the experimental data and performed the numerical simulations. A.G.G. participated in the fitting and the simulation. C.M. participated in the THz s-SNOM experiments. A.C. proposed the concept and developed the method of FIB fabrication of ultra-sharp metal tips and recorded the TEM image. A.B. helped with identifying and analyzing the topography-free test sample. All authors discussed the results. R.H. supervised the work. S.M., A.A.G. and R.H wrote the manuscript with input from all other co-authors.

**Acknowledgements**

The authors would like to thank Christopher Tollan (CIC Nanogune, San Sebastián, Spain) for the preparation of a lamella cross section of the sample used in this work, as well as Ken Wood and QMC Instruments Ltd. (Cardiff, UK) for providing the bolometer for THz detection. The authors acknowledge support from the Spanish Ministry of Economy, Industry, and Competitiveness (national project MAT2015-65525 and the project MDM-2016-0618 of the Marie de Maeztu Units of Excellence Program), the H2020 FET OPEN project PETER (GA#767227) and the Swiss National Science Foundation (Grant No. 172218).

**Bibliography**

(1) F. Keilmann und R. Hillenbrand, Nano-Optics and Near-Field Optical Microscopy, A. V. Zayats und D. Richards, Hrsg., Boston/London: Artech House, 2009.

(2) J. M. Atkin, S. Berweger, A. C. Jones und M. B. Raschke, „Nano-optical imaging and spectroscopy of order, phases, and domains in complex solids," *Advances in Physics,* Bd. 61, Nr. 6, pp. 745-842, 2012.

(3) E. A. Muller, B. Benjamin Pollard und M. B. Raschke, „Infrared Chemical Nano-Imaging: Accessing Structure, Coupling, and Dynamics on Molecular Length Scales," *The Journal of Physical Chemistry Letters,* Bd. 6, pp. 1275-1284, 2015.

(4) S. Amarie, P. Zaslansky, Y. Kajihara, E. Griesshaber, W. W. Schmahl und F. Keilmann, „Nano-FTIR chemical mapping of minerals in biological materials," *Beilstein Journal of Nanotechnology,* Bd. 3, pp. 312-323, 2012.

(5) I. Amenabar, S. Poly, W. Nuansing, E. H. Hubrich, A. A. Govyadinov, F. Huth, R. Krutokhvostov, L. Zhang, M. Knez, J. Heberle, A. M. Bittner und R. Hillenbrand, „Structural analysis and mapping of individual protein complexes by infrared nanospectroscopy," *Nature Communications,* Bd. 4, Nr. 2890, 2013.

(6) J.-J. Greffet und R. Carminati, „Image Formation in Near-field Optics," *Progress in Surface Science,* Bd. 56, Nr. 3, pp. 133-237, 1997.

(7) M. B. Raschke und C. Lienau, „Apertureless near-field optical microscopy: Tip-sample coupling in elastic light scattering," *Applied Physics Letters,* Bd. 83, Nr. 24, pp. 5089-5093, 2003.

(8) K.-T. Lin, S. Komiyama und Y. Kajihara, „Tip size dependence of passive near-field microscopy," *Optics Letters,* Bd. 41, Nr. 3, pp. 484-487, 2016.

(9) G. D. Boreman, Modulation Transfer Function in Optical and Electro-Optical Systems, Bellingham, Washington: SPIE - The International Society for Optical Engineering, 2001.

(10) C. S. Williams und O. A. Becklund, Introduction to the Optical Transfer Function, Bellingham, Washington: SPIE - The International Society for Optical Engineering, 2002.

(11) W. J. Smith, Modern Optical Engineering, New York City, New York: McGraw Hill, 2000.

(12) M. Born und E. Wolf, Principles of Optics, Cambridge: Cambridge Univercity Press, 1999.

(13) R. Hillenbrand und F. Keilmann, „Material-specific mapping of metal/semiconductor/dielectric nanosystems at 10 nm resolution by backscattering near-field optical microscopy," *Applied Physics Letters,* Bd. 80, Nr. 1, pp. 25-27, 2002.

(14) M. B. Raschke, L. Molina, T. Elsaesser, D. H. Kim, W. Knoll und K. Hinrichs, „Apertureless Near-Field Vibrational Imaging of Block-Copolymer Nanostructures with Ultrathin Spatial Resolution," *ChemPhysChem,* Bd. 6, Nr. 10, pp. 2197-2203, 2005.

(15) A. J. Huber, F. Keilmann, J. Wittborn, J. Aizpurua und R. Hillenbrand, „Terahertz Near-Field Nanoscopy of Mobile Carriers in Single Seminconductor Nanodevices," *Nano Letters,* Bd. 8, Nr. 11, pp. 3766-3770, 2008.

(16) K. Moon, Y. Do, M. Lim, G. Lee, H. Kang, K.-S. Park und H. Han, „Quantitative coherent scattering spectra in apertureless terahertz pulse near-field microscopes," *Applied Physics Letters,* Bd. 101, Nr. 011109, 2012.

(17) P. Dean, O. Mitrofanov, J. K. I. Keeley, L. Li, E. H. Linfield und A. G. Davies, „Apertureless near-field terahertz imaging using the self-mixing effect in a quantum cascade laser," *Applied Physics Letters,* Bd. 108, Nr. 9, p. 091113, 2016.

(18) K. Moon, H. Park, J. Kim, Y. Do, S. Lee, G. Lee, H. Kang und H. Han, „Subsurface Nanoimaging by Broadband Terahertz Pulse Near-field Microscopy," *Nano Letters,* Bd. 15, pp. 549-552, 2012.

(19) B. Hecht, H. Bielefeldt, Y. Inouye, D. W. Pohl und L. Novotny, „Facts and artifacts in near-field optical microscopy," *Journal of Applied Physics,* Bd. 81, p. 2492, 1997.

(20) T. Kalkbrenner, M. Graf, C. Durkan, J. Mlynek und V. Sandoghdar, „High-contrast topography-free sample for near-field optical microscopy," *Applied Physics Letters,* Bd. 76, Nr. 9, p. 1206, 2000.

(21) T. Taubner, R. Hillenbrand und F. Keilmann, „Performance of visible and mid-infrared scattering-type near-field optical microscopes," *Journal of Microscopy,* Bd. 210, Nr. 3, pp. 311-314, 2003.

(22) V. E. Babicheva, S. Gamage, M. I. Stockman und Y. Abate, „Near-field edge fringes at sharp material boundaries," *Optics Express,* Bd. 25, Nr. 20, pp. 23935-23944, 2017.

(23) M. Schnell, P. S. Carney und R. Hillenbrand, „Synthetic optical holography for rapid nanoimaging," *Nature Comm.,* Bd. 5, Nr. 3499, 2014.

(24) *Personal Communication*.

(25) S. Mastel, M. B. Lundeberg, P. Alonso-González, G. Yuando, K. Watanabe, T. Taniguchi, J. Hone, F. H. L. Koppens, A. Y. Nikitin und R. Hillenbrand, „Terahertz Nanofocusing with Cantilevered Terahertz-Resonant Antenna Tips," *Nano Letters,* Bd. 17, Nr. 11, pp. 6526-6533, 2017.

(26) C. Liewald, S. Mastel, J. L. Hesler, A. J. Huber, R. Hillenbrand und F. Keilmann, „All-electronic terahertz nanoscopy," *Optica,* Bd. 5, Nr. 2, pp. 159-163, 2018.

(27) J. N. Walford, J. A. Porto, R. Carminati, J.-J. Greffet, P. M. Adam, S. Hudlet, J.-L. Bijeon, A. Stashkevich und P. Royer, „Influence of tip modulation on image formation in scanning near-field optical microscopy," *Journal of Applied Physics,* Bd. 89, Nr. 9, pp. 5159-5169, 2001.

(28) R. Krutokhvostov, A. A. Govyadinov, J. M. Stiegler, F. Huth, A. Chuvilin, P. S. Carney und R. Hillenbrand, „Enhanced resolution in subsurface near-field optical microscopy," *Optics Express,* Bd. 20, Nr. 1, pp. 593-600, 2012.

(29) A. A. Govyadinov, S. Mastel, F. Golmar, A. Chuvilin, P. S. Carney und R. Hillenbrand, „Recovery of Permittivity and Depth from Near-Field Data as a Step toward Infrared Nanotomography," *ACS Nano,* Bd. 8, Nr. 7, p. 6911–6921, June 2014.

(30) B. Knoll und F. Keilmann, „Enhanced dielectric contrast in scattering-type scanning near-field optical microscopy," *Optics Communications,* Bd. 182, pp. 321-328, 2000.

(31) J.-L. Bijeon, P.-M. Adam, D. Barchiesi und P. Royer, „Definition of a simple resolution criterion in an Apertureless Scanning Near-Field Optical Microscope (A-SNOM): contribution of the tip vibration and lock-in detection," *The European Physical Journal Applied Physics,* Bd. 52, pp. 45-52, 2004.

(32) F. Huth, A. Chuvilin, M. Schnell, I. Amenabar, R. Krutokhovstov, S. Lopatin und R. Hillenbrand, „Resonant Antenna Probes for Tip-Enhanced Infrared Near-Field Microscopy," *Nano Letters,* Bd. 13, pp. 1065-1072, 2013.

(33) A. Wang und M. J. Butte, „Customized atomic force microscopy probe by focused-ion-beam-assisted tip transfer," *Applied Physics Letters,* Bd. 105, Nr. 053101, 2014.

(34) R. Esteban, R. Vogelgesang und K. Kern, „Full simulations of the apertureless scanning near field optical microscopy signal: achievable resolution and contrast," *Optics Express,* Bd. 17, Nr. 4, pp. 2518-2529, 2009.

(35) T. Taubner, F. Keilmann und R. Hillenbrand, „Nanoscale-resolved subsurface imaging by scattering-type near-field optical microscopy," *Optics Express,* Bd. 13, Nr. 22, pp. 8893-8899, 2005.

(36) E. D. Palik, Handbook of optical constants of solids II, London: Academic Press, 1991.

(37) J. D. Jackson, Classical Electrodynamic, Hoboken: John Wiley and Sons, Inc., 1999.

(38) F. Lu, J. Mingzhou und M. A. Belkin, „Tip-enhanced infrared nanospectroscopy via molecular expansion force detection," *Nature Photonics,* Bd. 8, Nr. 4, pp. 307-312, 2014.

(39) I. Rajapaksa, K. Uenal und H. K. Wickramasinghe, „Image force microscopy of molecular resonance: A microscope principle," *Applied Physics Letters,* Bd. 97, Nr. 073121, 2010.